\documentclass{iau} 
\usepackage{graphicx}

\title[IAU 363. High field NSs in binaries] 
{High magnetic field neutron stars and magnetars in binary systems}

\author[Sergei B. Popov]   
{Sergei B. Popov$^1$}

\affiliation{$^1$Sternberg Astronomical Institute, Lomonosov Moscow State University, \\ Universitetski pr. 13,
119234, Moscow, Russia \\ email: {\tt polar@sai.msu.ru}}

\pubyear{2022}
\volume{363}  
\jname{Neutron Star Astrophysics at the Crossroads:
Magnetars and the Multimessenger Revolution}
\editors{E. Troja \& M. Baring, eds.}
\begin{document}

\maketitle

\begin{abstract}
 Situation with highly magnetized neutron stars in binary systems is not yet certain. On the one hand, all best studied magnetars seem to be isolated objects. On the other, there are many claims based on model-dependent analysis of spin properties or/and luminosity of neutron stars in X-ray binaries in favour of large fields. In addition, there are a few results suggesting a magnetar-like activity of neutron stars in close binary systems. Most of theoretical considerations do not favour even existence, not speaking about active decay, of magnetar-scale fields in neutron stars older than $\sim10^6$~yrs.  However, alternative scenarios of the field evolution exist. I provide a brief review of theoretical and observational results related to the presence of neutron stars with large magnetic field in binaries and discuss perspectives of future studies. 
\end{abstract}

\firstsection 
\section{Introduction}

 What do we call a magnetar: just a neutron star (NS) with a large, $\gtrsim10^{14}$~G, magnetic field, or a NS with a detected specific activity powered by the field energy, $E_\mathrm{mag}$? Below I will distinguish these two cases, calling the first as high-B NSs and only the second as magnetars. Jointly, I call these classes as highly magnetized NSs. Of course, it well can be that a magnetar is just an active phase of the NS with large $E_\mathrm{mag}$, i.e. the same object can be observed as a high-B NS and as a magnetar at different time. If we focus on two important theoretical questions --- how to form a large magnetic field and how to save it for relatively long time, --- then the difference between two classes is not that important. If we choose an observational approach, then the difference is significant. Either we look for a magnetar-like activity of NSs in binaries (for example, X-ray systems) and search for components of already known magnetars, or we use various methods to estimate magnetic field of NSs in binaries aiming at values $\gtrsim10^{14}$~G. In this brief review I discuss all these approaches and questions in order to draw with broad strokes a picture presenting NSs with large magnetic fields in binary systems. 
 
 All well-established and well-studied magnetars --- anomalous X-ray pulsars (AXPs) and soft gamma-ray repeaters (SGRs),--- are isolated objects (see, however, Chrimes et al. contribution, this volume). In the next section, at first, I discuss evolutionary scenarios which can explain this feature. To continue theoretical considerations, I briefly sketch models of field decay which generally predict that at typical ages of NSs in low-mass X-ray binaries (LMXBs), and even in high-mass X-ray binaries (HMXBs), fields are expected to be lower than the magnetar values, or at least $E_\mathrm{mag}$ is not actively dissipated. In Sec.~3  several observation-based arguments in favour of presence of high-B NSs and even magnetars in known binaries are demonstrated. Then in Sec.~4  a scenario to save fields up to $10^{14}$~G at ages up to $\sim10$~Myrs is described. In the final section I present my conclusions. 

 There are several recent reviews closely related to the subject of this paper. 
 In the first place, I want to mention a brief review by \cite[Revnivtsev \& Mereghetti (2016)]{rm16} where many aspects of highly magnetized NSs in binary systems are touched, including some basics. Among many reviews on magnetars, I mention \cite[Turolla, Zane \& Watts (2015)]{tzw}. Finally, properties and evolution of interacting binaries are reviewed by \cite[Postnov \& Yungelson (2014)]{py14}.




\section{Do we expect magnetars in binaries?}

 An order of magnitude estimate is that $\sim10$\% of NSs are born as magnetars
 (see e.g., \cite[Kouveliotou et al. 1998]{kov98}, \cite[Popov et al. 2010]{petal2010}).
Population synthesis \cite[(Popov \& Prokhorov 2006)]{pp2006}, as well as observations \cite[(Kochanek 2021)]{kochanek21}, suggests that $\sim10$\% of NSs can stay in bound binaries after the first supernova (SN) explosion. Thus, neglecting all selection effects and/or correlations, one can expect that among circa 30 magnetars there might be three objects in binaries and among hundreds of known binary NSs we can find tens of magnetars. Of course, reality is more complicated. These simple estimates are no valid and we need to discuss more detailed models of binary evolution and magnetic field behavior. 

\subsection{Evolutionary considerations and the origin of magnetars}

 High-B NSs and magnetars might form a minority among compact objects. Thus, 
 it is reasonable to expect that they are formed through some specific channel(s) of stellar evolution. 
 
 Starting from the pioneering paper by \cite[Thompson \& Duncan (1993)]{dt93} (see also \cite[Bonanno et al. 2005]{bub05}) most popular ideas on the origin of high magnetic fields in NSs are related to dynamo mechanisms. Simulations show that rapid rotation is necessary to produce large dipole poloidal field (\cite[Raynaud et al. 2020]{ray20}). Progenitors of NSs can lose angular momentum via strong stellar winds, in addition they can significantly inflate on late evolutionary stages. If stellar cores (which later give birth to compact objects) are strongly coupled with envelopes, then it is difficult to expect very rapid rotation of NSs originated from isolated stars or in non-interacting binaries (see \cite[Postnov et al. 2016]{pos16}  on core-envelope coupling and e.g., \cite[Langer et al. 2008]{langer} and references therein on the role of rotation in massive star evolution, including binary evolution).
 
 Fast rotation of a stellar core can be a result of evolution in a binary system. There are three main routes to obtain a rapidly rotating core: angular momentum transfer via accretion, coalescence, and tidal synchronization. \cite[Popov \& Prokhorov (2006)]{pp2006} studied the role of the first two channels in production of highly magnetized NSs, and \cite[Bogomazov \& Popov (2009)]{bp2009} analysed the third one.

 In particular, \cite[Popov \& Prokhorov (2006)]{pp2006} demonstrated that evolutionary tracks resulting in spin-up of a NS progenitor due to accretion or coalescence (partly accounting for the effects of tidal synchronization) can on the one hand, explain $\gtrsim10$\% fraction of magnetars, and on the other hand, mostly produce isolated magnetars, in correspondence with observational data of that time. 
 
 In addition, binary evolution can open an alternative channel for formation of highly magnetized NSs. Coalescence of normal stars can result in enhanced magnetic field, as it was shown for the case of $\tau$ Boo
(\cite[Schneider et al. 2019]{schneider19}). If all of the magnetic flux in this star is conserved until core collapse of the merger product, then a resulting NS would have a surface magnetic field strength of about $10^{16}$~G. I.e., in this case a dynamo in a proto-NS is not necessary, just the magnetic flux conservation is enough to secure a magnetar-scale field (see also \cite[Makarenk, Igoshev \& Kholtygin 2021]{mik} and references therein on the role of fossil fields). 

 Despite some claims (see e.g., \cite[Davies et al. 2009]{df2009}, \cite[Clark et al. 2014]{clark14}), it is not certain if binary evolution plays a crucial role in formation of highly magnetized NSs.  Even if large fields are formed we face another effect which potentially can prevent existence of a significant number of highly magnetized NSs in not-so-young binaries --- the field decay.

\subsection{Magnetic field decay}

 The topic of magnetic field evolution was reviewed many times during last few years as this field is rapidly developing now (see e.g., \cite[Pons \& Vigano 2019]{pv19}, \cite[Igoshev et al. 2021]{iph21} and references therein). Here I just briefly summarize the main features, mostly following \cite[Cumming et al. (2004)]{cumming2004}.
 
 Dissipation of magnetic field occurs via two Ohmic processes: electron scattering off phonons and off crystalline impurities. We can define conductivities related to these two processes: $\sigma_\mathrm{ph}$ and $\sigma_\mathrm{Q}$. The total conductivity can be written as:
 
 \begin{equation}
\sigma = \frac{\sigma_\mathrm{ph}\sigma_\mathrm{Q}}
{\sigma_\mathrm{ph} +\sigma_\mathrm{Q}}.
\end{equation}

The corresponding Ohmic decay time scale is written as:

\begin{equation}
\tau_\mathrm{Ohm} = \frac{4\pi \sigma L^2}{c^2} .
\label{e:timescale_ohm}
\end{equation}
 Here $L$ characterizes the field length scale. Obviously, small-scale fields dissipate faster. 
 
 For strong fields there is a very effective (non-dissipative by itself) process which reduces the length scale --- the Hall cascade. It operates on the time scale:
 
 \begin{equation}
\tau_\mathrm{Hall} = \frac{4\pi e n_\mathrm{e}L^2}{cB(t)},
\end{equation}
where $n_\mathrm{e}$ is the electron number density and $e$ is the elementary charge. 
Note inverse dependence on the magnetic field. For fields $\sim 10^{15}$~G $\tau_\mathrm{Hall}$ can be as small as $\sim10^3$~yrs. 

Ohmic timescales can be written as (e.g., \cite[Igoshev et al. 2021]{iph21}): 

\begin{equation}
\tau_\mathrm{ph} \approx 80 \; \mathrm{Myr}\, \left(\frac{L}{1 \;\mathrm{km}}\right)^2 \left(\frac{\rho_{14}^{7/6}}{T^2_8}\right)\left(\frac{Y_e}{0.05}\right)^{5/3} ,
\end{equation}

and 

\begin{equation}
\tau_Q \approx 200 \; \mathrm{Myr}\, \left(\frac{L}{1 \;\mathrm{km}}\right)^2\left( \frac{\rho_{14}^{1/3}}{Q}\right)
\left(\frac{Y_e}{0.05}\right)^{1/3}\left( \frac{Z}{30}\right).     
\end{equation}
In these equations  $\rho_{14}$ is the density in units $10^{14}$~g~cm$^{-3}$, $T_8$ is temperature in the crust in units $10^8$~K, and $Y_e$ is the electron fraction in the layer where currents are mainly localized.
The parameter $Q$ shows how ordered is the crystalline structure:
\begin{equation}
Q=n_\mathrm{ion}^{-1}\Sigma_i\,n_i \times(Z^2-\langle Z\rangle^2), 
\end{equation}
where $Z$ is the ion charge, and $n$ is number density.

 If the Hall cascade effectively operates on a short time scale reducing $L$ down to small values, then Ohmic scales defined in eqs.~(2.4) and (2.5) become short. Thus, the decay time scale is regulated by eq.~(2.3). With the total amount of $E_\mathrm{mag}\sim 10^{47} B_{15}^2R_6^3$~erg and $\tau_\mathrm{Hall}\sim 10^3$~yrs, an average luminosity can be $\sim10^{35}$~--~$10^{36}$~erg~s$^{-1}$,
 in correspondence with magnetar observations. However, with such effective energy release the active life time of a magnetar is limited by a few initial Hall time scale.  After the field reaches a value $\sim10^{13}$~G the decay starts to be regulated by one of the Ohmic time scales. As a NS cools down, the time scale related to phonons starts to be very long, so decay is dominated by the processes associated with impurities in the crust, i.e. at this stage the driving parameter is $Q$. 
 
 However, in last several years numerical simulations of  field decay in young 
strongly magnetized NSs demonstrate that a rapid Hall cascade can be gradually terminated as the field configuration changes and approaches so-called Hall attractor (\cite[Gourgouliatos \& Cumming 2014]{gc14}). I will come back to this possibility in sec. 4. 

\section{Why do we need high-B NSs in binaries?}

{\it “All theory is gray, my friend. But forever green is the tree of life.” } (Goethe)\\
Despite general theoretical pessimism regarding significant number of highly magnetized NSs in e.g. HMXBs and LMXBs,  there are many claims based on, typically indirect, observational evidence. In this section I sketch the main examples.

\subsection{High luminosity sources}

 Discovery of the first ultra-luminous X-ray pulsar (PULX) by \cite[Bachetti et al. (2014)]{b14} initiated an active discussion about the origin of super-Eddington emission from NSs.\footnote{A recent list of PULXs can be found in the review by \cite[Fabrika et al. (2021)]{f21}.} 
 
 Accretion onto a NS with a large magnetic field might proceed via an accretion column formation. Properties of this feature depend on the field value (\cite[Basko \& Sunyaev 1975]{bs75}, \cite[Basko \& Sunyaev 1976]{bs76}). In particular, accretion luminosity can overcome the Eddington limit due to decreasing of the cross-section much below the Thompson value and due to geometrical effects (escape of radiation in the direction perpendicular to the column walls). In application to PULXs these ideas were pioneered by \cite[Mushtukov et al. (2015)]{m15}. For a recent development of this model see the contribution by V. Suleymanov et al., this volume. 
 
 Analysis of observational data on several PULXs  does not favour strong dipolar field. For example, studies of the PULX Swift J0243.6+6124 by \cite[Tsygankov et al. (2018)]{ts18} suggest an upper limit on the dipolar field $B\lesssim 10^{13}$~G basing on the absence of transition to the Propeller stage when the luminosity drops. Pulsed fraction modeling for this source is in reasonable correspondence with this limit. Thus, at least in some cases large dipole can be excluded for PULXs. 
 Then, it is possible to use advantages of high-field scenario applying it to higher order multipoles to explain super-Eddington luminosity and use weaker dipolar fields to fit spin properties of the NS. Such approach was used by \cite[Brice et al. (2021)]{brice21} (see also this volume). 
 
 The general idea is obvious. The dipolar component decreases much slower with distance ($B \propto r^{-3}$) in comparison with higher multipoles. Spin properties and transitions between different evolutionary stages depend on the field pressure at the magnetospheric boundary. Oppositely, properties of an accretion column are determined by the field value much close to the surface. In the latter case multipoles (in particular, an octupole --- in the study by \cite[Brice et al. 2021]{brice21}) are responsible for large X-ray flux. 
 
 The model presented by \cite[Brice et al. (2021)]{brice21} can reproduce properties of several PULXs with the dipole field on the surface $\sim10^{13}$~G and the octupole one --- $\sim10^{14}$~G. However, some simplifications were made by the authors, so further studies are necessary.  

\subsection{Spin-based field estimates}

 In the previous subsection I mentioned sources in which large dipolar field can be in contradiction with spin properties of NSs (I include here not just spins by themselves, but also period derivative, and parameters of the transition between Accretor and Propeller stages). Now we discuss the opposite case, when spin properties point towards magnetar-scale fields. 
 
 Measurements of a NS spin and its derivative (together with flux) opens many model-dependent possibilities to derive an estimate of the dipole component of the magnetic field (see e.g., \cite[Chashkina \& Popov 2012]{cp12}, \cite[Klus et al. 2014]{klus14}, \cite[Shi, Zhang \& Li 2015]{shi15} and references therein). 
All such approaches imply a particular model of accretion and matter-field interaction at the magnetospheric boundary. As we are not certain about these subjects, the estimates are not very secure and often contain contradictory results.

For example, a very simple field estimate can be obtained from the equilibrium period hypothesis. In this case it is assumed that spin-up and spin-down torques applied to a NS are equalized. 
For a disc accretion a standard approach provides the following value of the equilibrium period:

\begin{equation}
    P_\mathrm{eq, disc}= 3 B_{14}^{6/7}\dot M_{20}^{-3/7}\, {\mathrm s}. 
\end{equation}
Here $B$ is the polar dipole magnetic field and $\dot M$ --- the accretion rate in units $10^{20}$~g~s$^{-1}$. 
In the case of PULXs we obtain a magnetar-scale field (see \cite[Chen, Wang \& Tong 2021]{cwt21} for a list of PULXs with magnetic field estimates based on spin properties). 
A similar formula can be derived for a wind accretion. Then, for slow winds with velocity $\ll 10^8$~cm~s$^{-1}$ and observed spins about few hundred seconds even for sub-Eddington accretion rates it is possible to obtain a field $\gg10^{13}$~G. An X-ray pulsar GX 301-2 was many times discussed in this respect (some other recent examples and appropriate references can be found e.g. in \cite[Igoshev \& Popov 2018]{ip18} and \cite[Xu et al. 2021]{xu21}). 

Recently, a different model of low-rate wind accretion without disc formation was developed by
\cite[Shakura et al. (2013)]{shak13}. It was dubbed a settling accretion, as a wast low-density relatively hot envelope is formed around an accreting NS and matter flows to the compact object subsonically.
In \cite[Postnov et al. (2014)]{post14} the authors compare spin-up/spin-down regimes in different models with those in the settling accretion approach. In contrast to the quasi-spherical supersonic model, in the new model by Shakura et al. a NS can reach large spin periods even with a  standard magnetic field $B\sim 10^{12}$~--~$10^{13}$~G:
\begin{equation}
    B=0.24\times 10^{12} \left(\frac{P/100\, {\mathrm s}}{P_\mathrm{orb}/10\, \mathrm{days}}\right)^{11/12} \dot M_{16}^{1/3} \left(\frac{v}{10^8 \, \mathrm{cm} \, \mathrm{s}^{-1}} \right)^{-11/3}\, \mathrm{G}.
\end{equation}
Here $P$ and $P_\mathrm{orb}$ are spin and orbital period, correspondingly, and $v$ is relative velocity of the stellar wind matter relative to the compact object. 
Thus, with the settling accretion model it is possible to explain several sources without involving magnetar-scale fields. However, for NSs with long spin periods and  low stellar wind velocity even the settling accretion model can allow to obtain very large field values, as it was proposed in the case of 3A 1954+319 (\cite[Bozzo et al. 2022]{bozzo}).

 It is possible to obtain field estimates without the assumption of the equilibrium. 
 E.g., \cite[Shi, Zhang \& Li (2015)]{shi15}, together with calculations based on $P_\mathrm{eq}$, presented analysis with explicit dependence of the estimated field on the period derivative. It is illustrative, that in the disc accretion model applied to many Be/X-ray systems they obtained surface dipolar fields mostly above $10^{14}$~G, up to $\sim10^{16}$~G. These absolutely unrealistic estimates demonstrate strong model dependence of field estimates on the base of spin properties.

A more complicated example of model-dependent field estimate based on spin and orbital parameters was recently presented by \cite[Bachetti et al. (2021)]{b21}. These authors analyse new data on the first PULXs --- M-82 X2. 
An orbital decay is detected for M82 X-2. This allows to estimate the amount of matter transfered from the donor to the compact object. The obtained accretion rate is $\sim150$ times larger than the Eddington value. This makes any kind of beaming unnecessary. The final conclusion based on this findings, is that most probably the magnetic field of the NS is indeed high. 
However, this result was already criticized by \cite[King \& Lasota (2021)]{kl21}. 

Finally, an interesting case is presented in \cite[Yoneda et al. (2020)]{yo20}.
These authors analyse data on the gamma-ray binary LS 5039. {\it Suzaku} and {\it NuSTAR} observations allowed to detect a spin period, $P\sim 9$~s, and its derivative, $\dot P\sim 3 \times 10^{-10}$. The NS is not accreting. The magneto-dipole formula then provides a field estimate:

\begin{equation}
    B\sim 3 \times 10^{19} \sqrt{ P \dot P} \, \mathrm{G} \sim  10^{15} \, \mathrm{G}.
\end{equation}

It is interesting, that according to \cite[Yoneda et al. (2020)]{yo20} the compact object in LS 5039 can be not just a high-B NS, but a magnetar as the total luminosity cannot be explained by rotational energy losses or accretion. Then, another reservoir is necessary: $E_\mathrm{mag}$. However, this consideration still does not provide a direct evidence in favour of a magnetar-scale field. Luckily, there is a possibility to obtain a better estimate which I discuss in the next subsection.
  
\subsection{Cyclotron lines from accreting NSs}

Observation of cyclotron lines is, may be, the best way to measure magnetic fields of NSs. It is valid for accreting compact objects (see the seminal paper by \cite[Tr\"umper et al. 1978]{trump}) and for isolated NSs, including magnetars (see e.g., \cite[Borghese et al. 2017]{borg17}, and references therein).

 Two types of cyclotron lines are expected: electron and proton lines.
 In the first case, the energy is $E_\mathrm{c,e}= 11.6\, B_{12}(1+z)^{-1}$~keV.
 In the second, $E_\mathrm{p,c}=0.63\, B_{14}(1+z)^{-1}$~keV. In both cases $z$ is the gravitational redshift. Note, that up to relatively recent time it was possible to detect lines only in the range from few tenths of keV up to $\sim 10$~keV. Now  {\it NuSTAR} allows to detect lines up to few tens of keV. Thus, for high magnetic fields  only proton cyclotron lines can be observed. Oppositely, for standard fields $\sim 10^{11}$~--~$10^{12}$~G --- only electron lines. 

\cite[Brightman et al. (2018)]{br18}  detected a 4.5~keV line from a PULX in the galaxy M51. If the line is due to the electron cyclotron resonance --- then the surface field is $\sim 6\times 10^{11}$~G. In the case of the proton line --- $B\sim 7\times 10^{14}$~G. The authors suggest that the latter case is more probable due to narrowness of the detected spectral feature. This is a strong argument in favour of a high-B NS in the accreting binary system.

\cite[Walton et al. (2018)]{w18} studied a PULX in the galaxy NGC~300.
The detected line has energy $\sim12$~--~13~keV, and the authors interpret it as an electron cyclotron line. Thus, the field has a standard value $\sim10^{12}$~G. 

\subsection{FRBs and binaries}

 Quite unexpectedly, new arguments in favour of magnetars in binary systems appeared thanks to studies of fast radio bursts (FRBs). 
 This phenomenon was discovered by \cite[Lorimer et al. (2007)]{lor7}. 
 Up to now numerous millisecond long radio bursts have been observed by different radio telescopes at frequencies from $\sim100$~MHz up to $\sim10$~GHz (see a review in 
 \cite[Zhang 2020]{z20}). Typically (in $\gtrsim 90$\% cases), we see just one flare from a given source. However, there are about few tens sources of repeating bursts, and from some of them hundreds or even thousands events are detected, already. 
 
 It was rapidly demonstrated that sources are at extragalactic distances. First indications were based solely on large dispersion measure (e.g. \cite[Lorimer et al. 2007]{lor7}). However, now for $\sim20$ FRBs  host galaxies are securely identified \footnote{See the on-line catalogue of FRB hosts at http://frbhosts.org/ (\cite[Heintz et al. 2020)]{frbhost}.}
 
 Already in 2007 it was proposed that FRBs can be due to giant flares of magnetars (\cite[Popov \& Postnov 2007]{pp07}). In 2020 this hypothesis got strong support from observations of simultaneous flares in radio and X/$\gamma$-rays from a known Galactic magnetar SGR1935+2154 (\cite[CHIME/FRB Collaboration 2020]{chime}, \cite[Bochenek et al. 2020]{boch}, \cite[Mereghetti et al. 2020]{mer}, \cite[Li et al. 2021]{li}, \cite[Ridnaia et al. 2021]{rid},  \cite[Tavani et al. 2021]{tav}). Now, different scenarios involving magnetars are considered as the most promising approach to explain properties of FRBs (\cite[Zhang 2020]{z20}).
 
 For a relatively long time no periodicity was detected in the FRB observational data. Finally, for two repeating sources of FRBs --- FRB 180916.J0158+65 and FRB121102, --- a specific type of periodicity was identified (\cite[CHIME/FRB collaboration 2020b]{chimeb}, \cite[Rajwade et al. 2020]{rw21}, \cite[Cruces et al. 2021]{cruces}). Bursts of these two sources are grouped in time. Episodes of enhanced activity repeat with the period $\sim 16$ days in the case of FRB 180916.J0158+65 and with an order of magnitude longer period --- for FRB 121102. 
 The origin of these periodic behaviour is not known, yet. Two obvious explanations, which fit the magnetar model of FRBs, were proposed: orbital modulation in a binary system (e.g., \cite[Lyutikov, Barkov \& Giannios 2020]{lbg}) and precession (\cite[Levin, Beloborodov \&  Bransgrove 2020]{lbb}).
 On the one hand, it is possible to bring parameters of hypothetical binaries with repeating FRB sources with the binary evolution models (see sec. 2.1) resulting in a magnetar formation from a stellar core with enhanced rotation (\cite[Popov 2020]{p20}). On the other hand, extreme FRB sources producing many bursts or/and sources in galaxies with very low starformation rate, might be produced not in a usual stellar core collapse, but via coalescence of compact objects (white dwarfs or/and NSs). In this case, precession seems to be a more realistic option. 
 Still, I want to underline that the question of magnetar existence in binaries is now linked to FRB studies, especially those with observed periodic activity. 
 
\subsection{Magnetar-like bursts in binaries}

 A smoking gun of magnetar presence in binaries would be a direct detection of X/$\gamma$ bursts. There are several claims of such events which I briefly review in this subsection.
 
 The first, and may be the best, example is given in \cite[Torres et al. (2012)]{torres}. With {\it Swift}-BAT these authors detected a magnetar-like X-ray (15-50 keV) burst from a well-known gamma-ray binary LS I +61 303. Duration of the burst is less than one second, and the spectrum looks thermal with $kT\sim 7.5$~keV. These properties are similar to some magnetar bursts. Luminosity of the burst was estimated as $\sim 2\times 10^{37}$~erg~s$^{-1}$, somehow lower than typical SGR weak flares, but more similar to the weakest AXP  bursts.  The authors propose that in the case of LS I +61 303 we are dealing with a magnetar with the field $B\lesssim 10^{14}$~G.
 
 Another example is more recent. SGR 0755-2933 was discovered due to a single burst in 2016, and a soft X-ray counterpart was proposed (\cite[Barthelmy et al. 2016]{bart}).
 \cite[Doroshenko et al. (2021)]{doro21} 
discuss  the nature of the counterpart (2SXPS J075542.5-293353) and conclude that it is an HMXB system. Despite Doroshenko et al. suggest that most probably SGR 0755-2933  and 2SXPS J075542.5-293353 are unrelated to each other, there is still the possibility (mentioned, of course, in \cite[Doroshenko et al. 2021]{doro21}) that the burst was produced by a NS in the HMXB. Further studies of this system are very much welcomed.

\section{How to make a high-B NS in a binary?}

In sec. 2.2 it was shown that magnetic field decay in the case of magnetars operates on the time scale $\lesssim10^4$~yrs, i.e. much shorter than typical ages of known NSs in binary systems. 
In this section, following \cite[Igoshev \& Popov (2018)]{ip18} I present a scenario which allows existence of high-B NSs with ages $\sim10$~Myrs.

Two processes with short time scales operate in a young magnetar: Hall cascade and Ohmic decay due to scattering off phonons. We have to switch off both. The former can be terminated if the Hall attractor is reached, the second one --- if the NSs cools down rapidly. As we are interested in forming relatively mature highly magnetized NSs, we have to reduce the resistivity due to crystalline impurities. This goal can be reached if the parameter $Q$, introduced in sec. 2.2 is as small as possible. 

 The Hall attractor is an absolutely necessary ingredient, otherwise the field significantly decays even for low $Q$ and cold NSs. This feature of the field evolution was introduced by \cite[Gourgouliatos \& Cumming (2014)]{gc14}. Calculations performed by these authors demonstrated that the attractor is reached after a few initial Hall time scales. E.g., for a NS with  $B_0\sim10^{14}$~G the field reaches a stable configuration after approximately a few hundred thousand years. Note, that the Hall cascade  is switched off earlier for larger magnetic fields.

Inclusion of the Hall attractor stage, which is switched on at the time equal to three initial Hall scales ($t=3\, \tau_\mathrm{Hall,0}$),  allows us to obtain a field $\sim 10^{14}$~G at ages $\sim$~few~Myrs for the initial field $B_0\sim10^{15}$~G. For $B_0\sim10^{16}$~G we obtain $B\sim10^{14}$~G at 10 Myrs.
In all these calculations we use $Q=1$.
This is enough to be in correspondence with the existence of high-B NSs in HMXBs. 
Accretion potentially can slightly reduce the magnetic field. But as the accretion rate is not high in HMXBs ($\dot M\sim 10^{16}$~g~s$^{-1}$) and duration of the stage is also not long, the NS can accrete just $\Delta M<10^{-3}\, M_\odot$. This value is not enough to modify the field value significantly (see e.g., \cite[Konar 2017]{konar}).

Note, that here we are not speaking about magnetars as by definition it is assumed that $E_\mathrm{mag}$ is not actively dissipating at the Hall attractor stage. However, one can speculate about brief periods of field instability which can result in an episodic  bursting activity. 

\section{Conclusions and perspectives}

 At the moment, we are not sure in what amount highly magnetized NSs are present in binary systems of different age. Properties of high-B NSs and magnetars in binaries are related to many areas of astrophysics of compact objects.
 
 The first group of questions concerns the origin of magnetars. We still do not know which stars produce magnetars, what fraction of these objects is due to more exotic processes than core collapse, are magnetic fields always significantly amplified at the proto-NS stage or in some cases flux conservation is sufficient to reach a magnetar-scale field, etc. Appearance of significant statistics of magnetars in binary systems might significantly advance our understanding of the origin of this extreme type of NSs. 
 
 No doubt, presence of mid-aged high-B NSs (or even magnetars) in binaries might influence models of magnetic field evolution, as presently we are not sure about scenarios which allow existence of NSs  with strong fields at ages $\gtrsim1$~Myr. In particular, studies of high-B NSs of a Myr age (and older) can help to verify the hypothesis of the Hall attractor.
 
 Robust identification of high-B NSs in X-ray binaries is very important for accretion theory. Presently, best studied accreting systems contain standard ($B\sim10^{12}$~G) or low field ($B\sim 10^9$~G) NSs. Accretion regimes for $B\gtrsim10^{14}$~G (especially, for low and moderate accretion rates) are not studied well enough, partly due to lack of strong motivation to address this complicated problem not having reliable known systems, partly due to impossibility to make detailed comparison with observations. Accretion flows in the case of strong fields can have interesting peculiarities. For example, NS fields $\gtrsim 10^{15}$~G can allow to form systems similar to polars, where a low-mass donor star is inside the magnetosphere of the compact object. 
 
 Finally, as magnetars are the main candidates to sources of FRBs --- it is very important to understand in which binaries they can appear and how this can influence properties of FRBs. 

\acknowledgements
 I thank dr. A. Mushtukov for useful comments and dr. A. Igoshev for a long-term collaboration, numerous discussions on the topic of the review, and comments on the manuscript. 
 Participation of the author in the Symposium was supported by the Russian Science Foundation, grant 21-12-00141.
 

\end{document}